\def\half{\tfrac{1}{2}}
\def\aaps{Astron. \& Astrophys. Supp.}
\def\apj{Astrophys. J. }

\def\apjl{Astrophys. J.  }
\def\apss{Ap. Space Sci. }

\def\aap{Astron. \& Astrophys. }
\def\mnras{Mon. Not. R. Astr. Soc. }

\def\Ables68{1968PASAu...1..172A}
\def\Accorsi01{2001NIMPA.474..273A}
\def\Band07{2007arXiv0710.4602B}
\def\Barthelmy05{2005SSRv..120..143B}
\def\Bouchet01{2001ApJ...548..990B}
\def\Busbooma97{1997JOSAA..14.1058B}
\def\Busboomb01{2001ApOpt..40.3894B}
\def\Byarda92{1992ExA.....2..227B}
\def\Byardb92{1992NIMPA.322...97B}
\def\Carolia84{1984NCimC...7..786C}
\def\Caroli87{1987SSRv...45..349C}
\def\Cash79{1979ApJ...228..939C}
\def\Cook84{1985ICRC....3..295F}
\def\Dicke68{1968ApJ...153L.101D}
\def\Fenimore78{1978ApOpt..17..337F}
\def\Fenimoreb78{1978ApOpt..17.3562F}
\def\Giles81{1981ApOpt..20.3068G}
\def\Gottesmana86{1986ITNS...33..745G}
\def\Gottesmanb89{1989ApOpt..28.4344G}
\def\Gourlay84{1984ApOpt..23.4111G}
\def\Grindlay95{1995ihea.conf..213G}
\def\Grindlayb04{2004SPIE.5168..402G}
\def\Gunson76{1976MNRAS.177..485G}
\def\Hammersley84{ 1984NIMPA.221...45H}
\def\Jensen03{2003A&A...411L...7J}
\def\Kopilovitch94{1994MNRAS.266..357K}
\def\intZand94{1994A&A...288..665I}
\def\Jager97{1997A&AS..125..557J}
\def\Levine96{1996ApJ...469L..33L}
\def\Lukea97{1997ApOpt..36.6612L}
\def\Lukeb98{1998ApOpt..37..856L}
\def\Markwardt05{2005ApJ...633L..77M}
\def\McConnell87{1987ICRC....2..309M}
\def\Mertza{1965trop.book.....M}
\def\Mertzb{1989SPIE.1159...14M}
\def\Oda83{1983AdSpR...2..207O}
\def\Palmieri74{1974Ap&SS..26..431P}
\def\Proctor79{1979MNRAS.187..633P}
\def\Rideout96{1996A&AS..120..579R}
\def\Sims85{1985NIMPA.228..512S}
\def\Sodin95{1995AstL...21..423S}
\def\Strong03{2003A&A...411L.127S}
\def\Schaefera03{2003NIMPA.500..263S}
\def\Schaeferb04{2004astro.ph..7286S}
\def\Skinnera95{1995ExA.....6....1S}
\def\Skinnerb03{2003A&A...411L.123S}
\def\Skinnerc94{1994MNRAS.267..518S}
\def\Tokanai96{1996SPIE.2808..563T}
\def\Vanderspek99{1999A&AS..138..565V}
\def\Vigneau03{2003SPIE.4851.1326V}
\def\Wild83{1983OptL....8..247W}
\def\Willingalea81{1981MNRAS.194..359W}
\def\Willingale84{1984NIMPA.221...60W}
\def\Willmore92{1992MNRAS.258..621W}
\def\Winkler03{2003A&A...411L...1W}

\documentclass[10pt,letterpaper,twocolumn]{article} 

\usepackage{ol2}
\usepackage[draft]{hyperref}
\usepackage{amsmath}

\begin{document}

\twocolumn[ 

\title{The sensitivity of coded mask telescopes}

\author{Gerald K. Skinner}

\address {University of Maryland, College Park, MD 20742, USA \\
                 \& CRESST  \& NASA GSFC, Greenbelt, MD 20771, USA}
\address{Corresponding author: skinner@milkyway.gsfc.nasa.gov}                 

 \begin{abstract}
 { Simple formulae are often used to estimate the sensitivity of coded mask X-ray or gamma-ray  telescopes, but these are strictly only applicable if a number of basic assumptions are met. Complications arise, for example, if a grid structure is used to support the mask elements, if the detector spatial resolution is not good enough to completely resolve all the detail in the shadow of the mask or if any of a number of other simplifying conditions are not fulfilled. We derive more general expressions for the Poisson-noise-limited sensitivity of astronomical telescopes using the coded mask technique,  noting explicitly in what circumstances they are applicable. 
The emphasis is on using nomenclature and techniques that result in simple and revealing results.
Where no convenient expression is available a procedure is given which allows the calculation of the sensitivity. We consider certain aspects of the optimization of the design of a coded mask telescope and show that when the detector spatial resolution and the mask to detector separation are fixed, the best source location accuracy is obtained when the mask elements are equal in size to the detector pixels.
  }
\end{abstract}


] 

\section{Introduction}

 Coded mask telescopes have been widely used in X-ray and gamma-ray astronomy, particularly at those energies where other imaging techniques are not available or where the wide field of view possible with the technique is important.   Recent examples of astronomical applications of the technique include the BAT instrument on the SWIFT spacecraft \cite{\Barthelmy05} and three of the instruments on INTEGRAL \cite{\Winkler03}; other examples are described in \cite{\Levine96,\Tokanai96, \Jager97, \Vanderspek99}.  The technique is based on  recording the shadow of a mask containing both transparent and opaque regions in a pattern that allows an image of the source of the radiation to be reconstructed. Coded mask imaging  has been reviewed by Caroli et al.  \cite{\Caroli87}, with more recent discussion of some aspects  of the technique by Skinner \cite{\Skinnera95}.
 
Many possible mask patterns have been discussed. The mask may simply  contain  randomly placed holes \cite{\Ables68,\Dicke68}  or it may be based on  geometric patterns \cite{\Mertza ,\Mertzb}  -- indeed almost any design can be used without losing the imaging capability \cite{\Skinnera95}.  Most work has  been based on  patterns comprising holes placed on a regular rectangular or hexagonal  grid  according to some algorithm. Discussion of the choice of algorithm  for placing the holes has concentrated on designs in which the (cyclic) autocorrelation function of the pattern, sampled at shifts corresponding to a whole number of cells of the grid, is bi-valued with a central peak and flat wings. An extensive literature   \cite{\Gunson76,\Fenimore78,\Proctor79,\Giles81,\Oda83,\Wild83,\Gourlay84,\Cook84,\Gottesmana86,\Byarda92,\Byardb92,\Kopilovitch94,
\Busbooma97,\Sodin95,\Lukea97,\Accorsi01} 
exists on  arrays which have this property, which are usually termed `Uniformly Redundant Arrays' (URAs).    For URA-based masks, in certain well defined circumstances, cross-correlation of the recorded data with an array which corresponds to the mask pattern (with a scaling and offset applied)  leads to images with a point source response function (PSF) having a central peak and perfectly flat side-lobes. As image reconstruction by cross-correlation  can be shown (again in specific circumstances, to be discussed below) also to yield the best possible signal-to-noise ratio, such solutions have attracted widespread attention. 

Variants  of URAs  have been proposed (e.g. Modified Uniformly Redundant Arrays, MURAs, \cite{\Gottesmanb89}; see also \cite{\Lukeb98})  in which the same ideal PSF is obtained when the reconstructing array differs marginally from the coding pattern. Provided the number of elements is large, the signal-to-noise ratio is essentially the same as for URAs.

URAs (and MURAs, etc) provide a mathematically satisfying solution to the problem of mask design, but their advantages in practice are less evident. The circumstances in which the ideal response is obtained relate to the cyclic nature of the patterns.  The PSF is free from spurious responses (`ghosts' or 'sidelobes') only if  the shadow recorded is always of a whole number of cycles of a repeating pattern. It can be arranged that this condition is met over a limited field  (the so-called `fully coded field of view', FCFOV)  but for sources outside this region (in the `partially coded field of view', PCFOV) the shadow of the edge of the mask will  appear in the recorded data. Sources in the PCFOV  can produce spurious responses  within the FCFOV and vice-versa.  It is possible to block with a  collimator consisting of slats or tubes the flux from sources in the PCFOV, but as pointed out by \cite{\Sims85} this not only narrows the observable field but results in attenuation of the recorded signal even within the FCFOV.  Thus it is often the case that random mask patterns are as good as any other and indeed a random pattern was selected for the mask of the very successful BAT instrument on the SWIFT satellite \cite{\Vigneau03}.

The  ghost images and other imaging artifacts that arise from partial coding, from non-uniform background, or from other effects present in real instruments can be alleviated provided one has  a good understanding of the instrument and adequate computing resources. Although  for sufficiently long observations or combinations of observations, these {\it systematic} errors will inevitably become important, many methods are available to reduce them \cite{\Hammersley84,\Willingalea81,\Willingale84,\McConnell87, \Willmore92,\Grindlay95,\Rideout96,\Bouchet01,\Skinnerb03,\Strong03,\Schaefera03 }.
Typically these involves fitting and subtracting bright sources or otherwise taking them into account, and carefully modeling background non-uniformities. 

Even with advanced image reconstruction techniques,  Poisson (or `photon') noise due to limited counting statistics leads to {\it random} errors, placing a limit on  the sensitivity. This limit is particularly important for very short observations of relatively bright sources, as is the case in the detection of gamma-ray bursts, for example. However  even for long term survey observations  it places an intrinsic limit on the sensitivity that can be achieved, however  sophisticated the analysis technique. 
 
 We here consider the calculation of the statistical limit to the significance with which a point source can be observed in the presence of Poisson noise on both the flux from the source and thpe detector background.  We pay  particular attention to the assumptions that are made in the derivation of  the formulae and the circumstances in which they are valid. In section \ref{assumptions_section} we list the assumptions that have been made, explicitly or implicitly, in many previous approaches to this problem. In successive sections we attempt to provide useful expressions for the signal-to-noise ratio where subsets of these assumptions do not hold.  
 
 As we deal only with the Poisson-noise-limited sensitivity of the instrument, the formulae given here only place an upper bound on the signal-to-noise ratio that can be obtained. The flux is assumed to be assessed by a procedure which maximizes the signal-to-noise-ratio for the source in the presence of an unknown uniform detector background. In effect, data from all the detector pixels are combined, with weights which achieve this objective, but which consequently do not necessarily minimize imaging artifacts.   Artifacts due to imperfect modeling of the background, to imaging in regions of the PCFOV where the instrument imaging response is intrinsically poor, or to using in the analysis an insufficiently precise  description of the instrument  response, may add systematic noise to the Poisson noise. If analysis methods are adopted that are designed to reduce such artifacts (for example by reducing or eliminating `ghost' responses  in images of fields  containing extended sources or multiple point sources), they may increase the effect of Poisson noise. In the limit inverse matrix techniques (or inverse filtering) may completely remove systematic errors but are well known to lead to noise amplification.
 
 By not considering errors other than statistical ones we effectively suppose that observations are short enough that systematic errors are well below the limit imposed by statistics and ignore the fact that for sufficiently long integration times they will eventually become important.
In a well designed instrument and with appropriate treatment of the data,  performance close to the Poisson limit can nevertheless be achieved even for comparatively long observations -- particularly if the telescope orientation is `dithered' or scanned  during the observation to reduce the systematic noise, as forms part of the INTEGRAL observing strategy \cite{\Winkler03}, as is  important for SWIFT/BAT survey work \cite{\Markwardt05}, and is planned for EXIST \cite{\Grindlayb04, \Band07}.    

It is emphasized  that the sensitivity considered is that for a point source at a known position or when measuring the flux in a particular  pixel of an image; any flux from other point sources, or from extended emission away from the pixel under consideration, is handled  by treating it as additional background.

 \section{Assumptions frequently made}
 \label{assumptions_section}
 
Simple analyses of the signal-to-noise ratio obtainable with a coded mask telescope  often assume, explicitly or implicitly, that the following conditions are met:
\begin{enumerate} 
\item  Half of the mask elements are open, half closed (mask element open fraction $f_e = {\frac{1}{2}}$).\label{a_50_50}
\item The holes are identical and equal in size to the pitch of the grid on which they are placed. For example, there is no supporting structure of the sort illustrated in Figure \ref{mask_fig} and the overall open fraction $f$ is then  equal to $f_e$. \label{a_no_grid}
\item The measurement uncertainty  is the same for every detector element. We here characterize the source  strength by the number of  counts per unit area of detector where the mask is open, $S$,  and the background by $B_{f,t}$, the number of background events  per unit area of detector. So  this implies  assuming that $B_{f,t}\gg S$. \label{a_bg_dominated} The subscript `${f,t}$' is a reminder that as the background generally contains a significant contribution from diffuse sky emission, $B_{f,t}$ will generally be a function of $f$, of the solid angle of the field of view, and of the transmission (discussed below) of the mask elements. It is often convenient to assume that  other sources in the field of view give rise to flux that is  uncorrelated with the shadow pattern corresponding to the source under consideration. Their contribution to the detector counts can then be  considered to be smeared out and included in $B_{f,t}$. Any such component of  $B_{f,t}$, too,  will depend on $f$ and the $t$ values.
\item Each mask element is either totally opaque ($t_0=0$)  or totally transparent ($t_1=1$). \label{a_no_leaks} If this condition is not met, and if there is a background component due to the sky or other sources,  then $B_{f,t}$ will also depend on the actual values of $t_0,t_1$. 
\item The number of events is such that Poisson statistics may be treated as Gaussian\label{a_gaussian}. It will be shown in \S\ref{stats} that this supposition is a good one except in the most extreme circumstances, so that even where it is formally still being made below, it may often be ignored. 
\item The detector has perfect spatial resolution so that the exact position of arrival of each photon is known, as opposed to having finite size pixels or a realistic continuous position readout with some measurement uncertainty.  \label{a_perfect_det} 
\newcounter{saveenum}
 \setcounter{saveenum}{\value{enumi}}
\end{enumerate}

The signal-to-noise ratio most often evaluated is the estimate of the intensity of the flux from a source, relative to the uncertainty in its measurement ($S/\sigma_s$ in the terminology used below). This can be different from the value relative to the noise in the absence of the source ($S/\sigma_I$ below).   
We will generally consider the former parameter because the latter can readily be obtained from the same formulae, but where relevant  we will note as an assumption that it is indeed $S/\sigma_s$ that is required:

\begin{enumerate}
\setcounter{enumi}{\value{saveenum}}
\item
The relevant  signal-to-noise ratio is the estimate of the intensity of the flux from a source  relative to the uncertainty in its measurement.  This assumption is never needed if assumption 3 is made, as the two signal-to-noise ratio estimates are then the same. \label{a_snr_src} 
 \setcounter{saveenum}{\value{enumi}}
\end{enumerate}
{\noindent There are two more simplifying assumptions that we will generally continue to make:}

\begin{enumerate}
\setcounter{enumi}{\value{saveenum}}
\item  The sensitivity to be discussed is a typical value over part or all of the region imaged and/or the mask elements through which radiation is received are sufficiently numerous that numbers based on the average open fraction may be used.  Results are then the same for (M)URA-based masks and random ones. \label{a_typical}

\item The measurement uncertainty due to background counting statistics is uniform across the detector plane. Sometimes a highly non-uniform background may be modelled and subtracted out. Even if the expectation level of the residual background is then everywhere zero, the random fluctuations can be more important in some regions than others, in which case this assumption is not valid.
\label{a_uniform_bg}
\end{enumerate}

An example  of when assumption \ref{a_typical} is important arises  when the detector plane consists of pixels that are on the same pitch as the mask elements or one that is a submultiple of it. If one considers  only source directions such that the detector pixels are either fully shadowed or fully illuminated, then the sensitivity is the same as if the detector had perfect spatial resolution. However in other directions the sensitivity  is up to a factor of 2 poorer. It is generally most useful to average out such effects. An exception arises if an observation, or  sequence of observations,  is planned such that, for a particular source selected for study, the shadow boundaries always fall  between detector pixels (e.g. the ``7 point hexagonal" pointings of the INTEGRAL/SPI instrument \cite{\Jensen03}).

Some aspects of a real system may invalidate several of these assumptions. For example at high energies masks are likely to be partially transparent, contrary to assumption \ref{a_no_leaks}, and the large thicknesses which are employed to minimize the consequent loss in sensitivity  mean that the apparent hole size and shape becomes a function of off-axis angle  and one no longer has the simple situation assumed in \ref{a_no_grid}.

%
\section {Relaxing conditions \ref{a_50_50},\ref{a_no_grid} and \ref{a_bg_dominated} -- allowing masks with arbitrary pattern and detector background not necessarily dominant}
\label{section_arb}

We will consider first a coded mask telescope in which the mask pattern is not necessarily 50\% open, 50\% closed (breaking assumption \ref{a_50_50}). Furthermore we suppose  that  the elements are not necessarily  simple squares or hexagons and the mask pattern may contain structures other than the elements themselves  (breaking assumption \ref{a_no_grid}), an important example being the presence of a  supporting grid as shown in Figure \ref{mask_fig}. At the same time we will consider the general case where it is not necessarily true that $S\ll B_{f,t}$, breaking \ref{a_bg_dominated}. This case, and the associated issue of the optimum open fraction of the mask when $S\not<B_{f,t}$, have been discussed in the literature \cite{\Dicke68,\Palmieri74,\Gunson76,\Fenimoreb78,\Carolia84,\intZand94,\Accorsi01}, but we here try to provide a unified approach and to correct some errors that have arisen.

\subsection{Signal-to-noise ratio}

\begin{figure}[tbp] 
   \centering
   \includegraphics[width=3.45in]{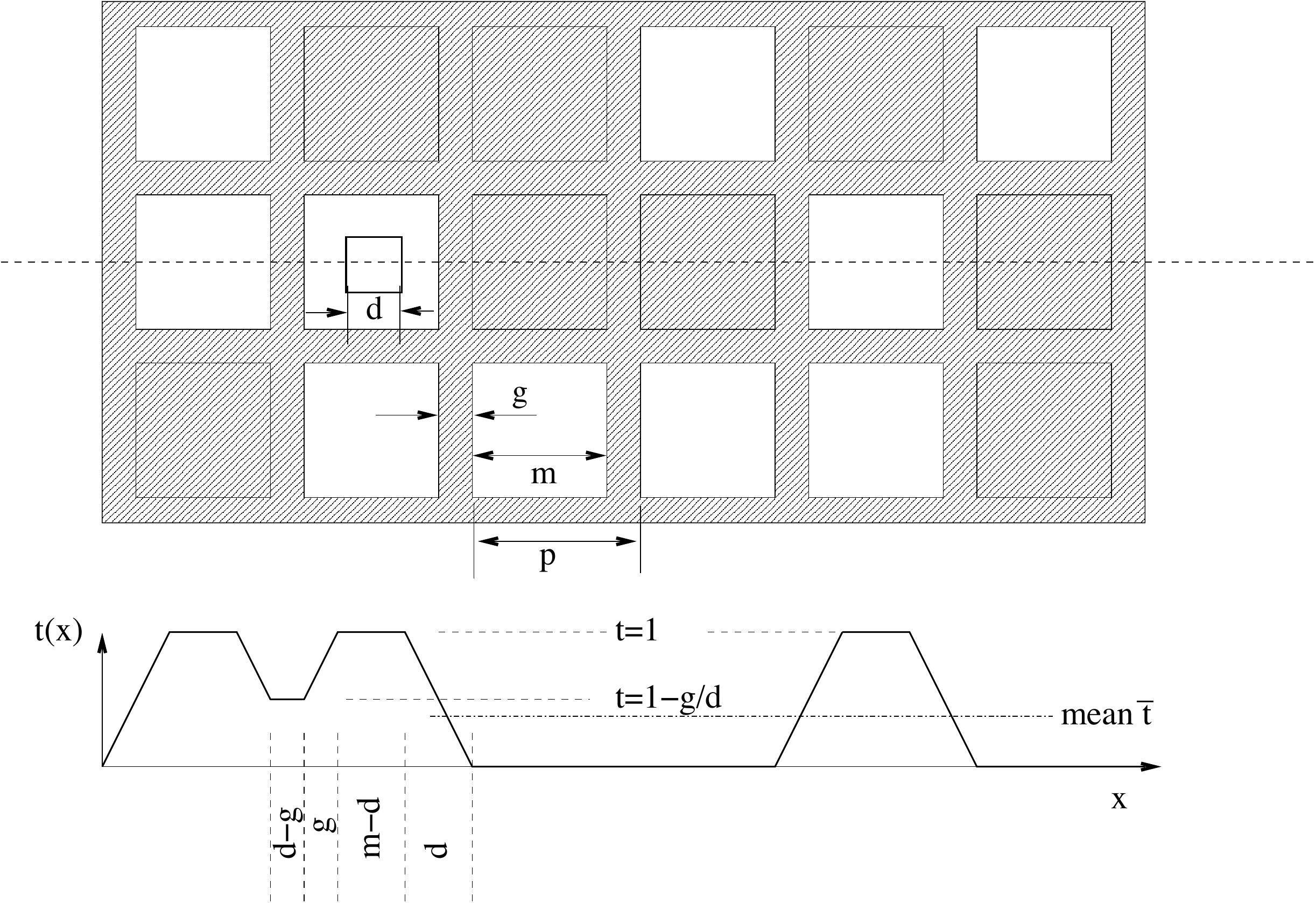} 
   \caption{A part of a mask in which the opaque elements are supported by a grid with bar width $g$ and pitch $p$, leaving holes of width $m$. The plot beneath shows the response of a square detector pixel of side $d$  as it is moved across the mask shadow along the line shown. The widths of the transition regions are indicated in the case where $g<d<m$. }
   \label{mask_fig}
\end{figure}

The important parameter in this case is the `open fraction' of the mask, $f$.  This takes into account the fraction of mask elements that are open, $f_e$, but may also be affected by other aspects of the design. Thus $f=f_e (m/p)^2$ in the case of the example structure in Figure \ref{mask_fig}.
As for the moment we still ignore any effects of finite detector resolution, the total detector area $A$ may be considered to be divided into an area $f A$  that sees the source plus detector background (cosmic and particle) and an area $(1-f)A$ that  just measures  background. If the source is in the PCFOV then $A$ should be taken as the area of that part of the detector that would, but for the mask, see the source. The expectation values for the counts measured in the two regions are 
\begin{equation}
C_S= f A (S+B_{f,t})
\label{eqn_c1}
\end{equation}
\begin{equation}
C_B= (1-f) A B_{f,t}
\label{eqn_c0}
\end{equation}

Our estimate of the source strength is then 
\begin{equation}
\hat S= {\frac{C_S}{f A}} - {\frac{C_B}{(1-f)A}} ,
\label{flux_estimate_eqn}
\end{equation}
with variance
\begin{equation}
\sigma_S^2= {\frac{C_S}{(f A)^2}} +{\frac{C_B}{(1-f)^2A^2}} \\
\label{var_eqn}
\end{equation}
\begin{equation}
= {\frac{S+B_{f,t}}{f A}} +{\frac{B_{f,t}}{(1-f)A}} .\\
\end{equation}

The signal-to-noise ratio of the source flux measurement is thus
\begin{equation}
{\frac{\hat S}{\sigma_S}}= S \sqrt{ \frac{f(1-f) A}{(1-f)S+B_{f,t}}}
\label{sn_basic_eqn}
\end{equation}
\hfill (assumptions  \ref{a_no_leaks}--\ref{a_uniform_bg}).\\

Note that in this case we obtain the same result whether $f$ deviates from $\half$ because  of a supporting grid as in Figure \ref{mask_fig}, or whether it  simply reflects the fraction of mask elements which are open ($f_e$) or  a combination of the two. Indeed it applies to an arbitrary mask design.

Some particular cases are  (a) the limiting case  $B_{f,t}\gg S$, for which
\begin{equation}
{\frac{\hat S}{\sigma_S}}= S \sqrt{ \frac{f(1-f) A}{B_{f,t}}}
\label{b_gt_s_eqn}
\end{equation}
\hfill (assumptions  \ref{a_bg_dominated}--\ref{a_uniform_bg})\\
and (b) the special case $f=\half$, for which the signal-to-noise ratio can be written
\begin{equation}
{\frac{\hat S}{\sigma_S}}= {\frac{(S/2)A}{ \sqrt {\left[(S/2)+B_{1/2}\right]A}}}=\left({\frac{S}{\sigma_S}}\right)_{ref}
\label{snr_ref_eqn}
\end{equation}
\hfill (assumptions  \ref{a_50_50},  \ref{a_no_leaks}--\ref{a_uniform_bg}),\\
which is simply the number of counts due to the source divided by the square root of all the counts (source plus background, the latter including any contribution from other sources in the field of view). Below we will use this  value as a reference against which to compare the sensitivity in other cases.

Finally (c) when both of these conditions apply, we have the widely quoted expression for the signal-to-noise ratio for an ideal 50\% open coded mask instrument in the background dominated case
\begin{equation}
{\frac{\hat S}{\sigma_S}}= {\frac{S}{2}} {\sqrt {\frac{A}{B_{1/2}}}}
\end{equation}
\hfill (assumptions  
\ref{a_50_50}, \ref{a_bg_dominated}--\ref{a_uniform_bg}).\\

The signal-to-noise ratio as defined above is the ratio of the source strength to the uncertainty in its measure. For knowing whether a source is significantly detected or not, a more appropriate measure is the ratio of the measured flux to the noise in the surrounding region of an image. Consider a test position away from the true source position. The expected distribution of events for a hypothetical source at this position should ideally be uncorrelated with the actual distribution due to the real source. If there is some residual correlation, then systematic effects (ghosts or sidelobes) will result. However we are here concerned with random noise so we may suppose that all the recorded events will be divided between the region measuring the flux from the hypothetical source and that measuring the background, in proportion with the areas of the two regions 
\begin{equation}
C_S'= f A (fS+B_{f,t}),
\end{equation}
\begin{equation}
C_B'= (1-f) A (fS + B_{f,t}).
\end{equation}
With these values Equation \ref{flux_estimate_eqn} gives an expectation value of zero\footnote{This \label{correl_note} illustrates an approximation in the approach used here. The shadow of a source at trial positions away from the peak cannot be {\bf totally} independent of that expected for a source at the peak position \cite{\Skinnerc94}. The process described here is equivalent to correlation with a mean-subtracted  form of the expected count distribution, which must produce a function with a mean of zero. Thus a positive peak implies a negative mean level elsewhere. The effect goes inversely with the number of resolution elements in the field of view and becomes negligible for sufficiently large $N$.}, with variance
\begin{equation}
\sigma_I^2= {\frac{1}{A(1-f)}}\left( {S}+{\frac{B_{f,t}}{f}}\right)
\end{equation}
leading to
\begin{equation}
{\frac{\hat S}{\sigma_I}}= S \sqrt{ \frac{f(1-f) A}{fS+B_{f,t}}}
\label{sn_rel_field}
\end{equation}
\hfill (assumptions 
\ref{a_no_leaks}--\ref{a_perfect_det}, \ref{a_typical}
, \ref{a_uniform_bg}),\\
which differs from Equation \ref{sn_basic_eqn} only in the factor multiplying $S$ and can be obtained from it by omitting that term and including the source flux in the background. The two are identical if $B_{f,t}\gg S$ or if $f=\half$. 

\subsection{Optimum choice of $f$}
\label{opt_f_sect}

If $B_{f,t}\gg S$, and if $B_{f,t}$ is independent of $S$, then  consideration of Equation \ref{b_gt_s_eqn} shows that the optimum open fraction is  50\%. However in general the background may not dominate. $B_{f,t}$ is the combination of a component intrinsic to the detector plus one due to a combination of diffuse sky emission and  the smeared effect of sources, other than the one of interest, in the field of view. Thus we may write $B_{f,t}=B_{det}+fB_{sky}$.  Putting $b=(B_{sky}/B_{det}) $ and $s =(S/B_{det})$
and solving for the  optimum value of $f$ one finds
\begin{equation}
   f_{opt}(b,s) =  \frac{ 1}{1+\sqrt{ \frac{1+b} {1+s}} } 
\label{opt_f_eqn}
\end{equation}
\hfill (assumptions  \ref{a_no_leaks}--\ref{a_uniform_bg}).\\
Equation \ref{opt_f_eqn} is equivalent to the expression given by in 't Zand et al.  \cite{\intZand94}. Although their result is correct, those authors state that it is the same as that of  Fenimore \cite{\Fenimoreb78}, which is in fact different.  The latter contains an additional factor of two, noted by  Accorsi et al. \cite{\Accorsi01}  as an error. It is in fact attributable to an attempt to combine the two different signal to noise ratios in Equations \ref{sn_basic_eqn} and \ref{sn_rel_field} above in a single expression, which in retrospect is probably not useful.

We can measure the advantage, $g$,  of using the optimum open fraction as the signal-to-noise ratio with $f=f_{opt}$, relative to that for $f=\half$ given by Equation \ref{snr_ref_eqn}.   After much manipulation it turns out that $g$  depends only on the value of $f_{opt}$ and is independent of the particular combination of $s$ and $b$ which led to that value. It is  simply given by 
\begin{equation}
   g(b,s)^2 = 1+4\left(f_{opt}(b,s)-{\frac{1}{2}}\right)^2.
   \label{advantage_eqn}  
 \end{equation}
\begin{figure}[tbp]
   \centering
   \includegraphics[width=3.4in,angle=0]{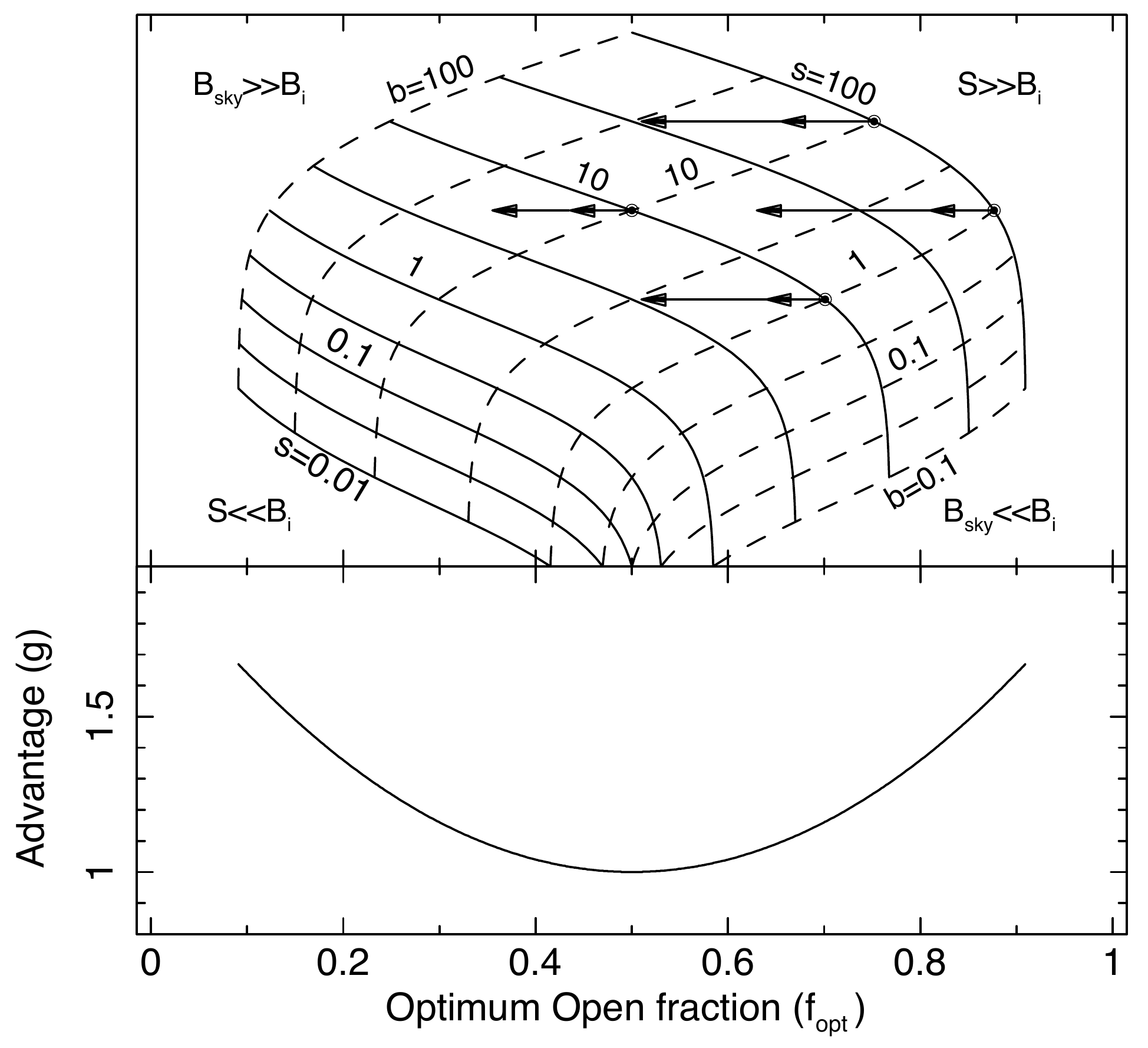} 
   \vspace{-5mm}
   \caption{Top:  Nomogram for determining the optimum mask open fraction $f_{opt}$ given the parameters $b$ and $s$, which are respectively  the sky background and the strength of the observed source, relative to the intrinsic detector background. $f_{opt}$ can be read from the horizontal scale at the intersection of lines of constant $s$ (continuous lines, logarithmically spaced) and of constant $b$ (dashed lines, also logarithmically spaced). The arrows illustrate the effect of a finite detector resolution (see \S  \ref{coding_power}\ref{fopt_detres}). Bottom: The signal-to-noise ratio when using $f = f_{opt}$ relative to that with $f=\half$. }
   \label{fopt_fig}
\end{figure}
Figure \ref{fopt_fig} provides a convenient nomogram for  $f_{opt}$ and $g$ which also illustrates some conclusions that can be drawn. One sees, for example, how large $b$ (strong background
 from the sky or from sources other than that of interest) favors low $f$, moving towards the single pin-hole camera extreme. On the other hand, for studying a bright source (large $s$)  high $f$, more like an open ``light bucket", are preferable. 
  
As has been noted by other authors, the advantage in signal-to-noise ratio to be obtained by using a value of $f$ other than $\half$ is small except in the most extreme circumstances. We note however that the low values of $f$ marginally favored  from this point of view when $S\ll B_{det}\ll B_{sky}$ can lead to important data handling and telemetry reductions, particularly when information about each event is recorded. 

If the source to be studied dominates over the effects of intrinsic background  by more than does the combination of all other sources and the diffuse sky emission ($s>b$), the optimum fraction can be larger than $\half$. The  circumstances in which this is most likely to be relevant is when the objective is to obtain information very quickly, for example when studying short bursts of emission. We note, however that this conclusion depends on the definition of signal-to-noise ratio.  

If it is the detectability of a source that is important, rather than the precision with which its intensity can be measured, then  it is ${\hat S}/{\sigma_I}$ (Equation \ref{sn_rel_field}) that should be optimized rather than ${\hat S}/{\sigma_S}$ (Equation \ref{sn_basic_eqn}).  The flux from the source should then be included in $B_{sky}$, not $S$, and Equations \ref{opt_f_eqn} and \ref{advantage_eqn} and Figure   \ref{fopt_fig}  used with $s$ set to zero (or a small value). Figure   \ref{fopt_fig}  shows that $f_{opt}$ is  always less than $\half$ in this case.

\section{Imperfect mask opacity/transparency : relaxing assumption  \ref{a_no_leaks}}

If the mask  elements are not perfectly opaque and transparent but have transmissions $t_0 and t_1$, respectively,  equations \ref{eqn_c1},\ref{eqn_c0} take the form 
\begin{eqnarray}
C_S&=& f A (t_1S+B_{f,t}) \\
C_B&= &(1-f) A (t_0S + B_{f,t}).
\end{eqnarray}
Note that $B_{f,t}$ will in this case be a function of $t_0,t_1$, as well as of $f$. Following the same logic as above one finds
\begin{equation}
{\frac{\hat S}{\sigma_S}}= S(t_1-t_0) \sqrt{  \frac{f(1-f) A}{\left[(1-f)t_1+ ft_0\right] S+B_{f,t}}}  
\label{basic_with_t}
\end{equation}
\hfill (assumptions  \ref{a_gaussian}--\ref{a_uniform_bg})\\%
\goodbreak
and for source detection
\begin{equation}
{\frac{\hat S}{\sigma_I}}= S(t_1-t_0) \sqrt{  \frac{f(1-f) A}{\left[(1-f)t_0+ ft_1\right] S+B_{f,t}}}
\label{sn_rel_field_with_t}
\end{equation}
\hfill (assumptions  
\ref{a_gaussian}, \ref{a_perfect_det}, \ref{a_typical}, \ref{a_uniform_bg}).\\

Thus the only changes necessary to allow for a uniform absorption in the nominally open areas  ($t_1<1$) and/or for uniform leakage through the closed ones ($t_0>0$) are to multiply the signal-to-noise ratio by a factor  $t_1-t_0$ and to correct the noise contribution due to source counts if this is not negligible. The reason for the simple form of the multiplying factor will become evident in Section \ref{coding_power}.

Accorsi et al. \cite{\Accorsi01} have treated the question of the optimum $f$ when the mask is leaky ($t_0>0$), but unfortunately the expression they give contains  some typographical errors, as well as being rather complex (although their equation for the signal-to-noise ratio is correct, that for $f_{opt}$ twice has $2t$ in place of $t$). The general case, $t_0>0$ and $t_1<1$, can, however,  be handled by using the equations and nomogram of \S\ref{section_arb}\ref{opt_f_sect}  with adjusted parameters  
\begin{eqnarray}
s' & = &t_1s+t_0b \nonumber \\
b' &=& t_0s +t_1b. 
\end{eqnarray}
in place of $s,b$.
 
\section{Gaussian or Poissonian statistics? : Assumption \ref{a_gaussian}  }
\label{stats}
In the above, the only respect in which it has been assumed that Gaussian statistics are applicable is in characterizing the signal-to-noise ratio and the significance of detection in terms of a standard deviation calculated as the root of the sum  of the variances (Equation \ref{var_eqn}, or its equivalents in other cases). If both $C_S$ and $C_B$ are small this is a slight simplification as the distribution of $\hat S$ will not be strictly Gaussian (or Poissonian).   The resulting effects are tiny  except where only a very few events in total are involved.
 A potentially relevant case arises in the detection of a very brief burst -- one occurring in so short a time that the background is very small. In this case the significance of detection should ideally be expressed in terms of likelihood. Often in these circumstances the background rate (and even its distribution over the detector) will be well determined by considering data before, and perhaps after, the burst. However to place the problem in the same context as the above discussion we consider the case where there is no information available outside the time of the event itself. For the same reason we will consider the significance of detection {\it at a given position}, without considering the degrees of freedom associated with finding the location of the event.

 \begin{figure}[tbp]
   \centering
   \includegraphics[width=3.4in,angle=0]{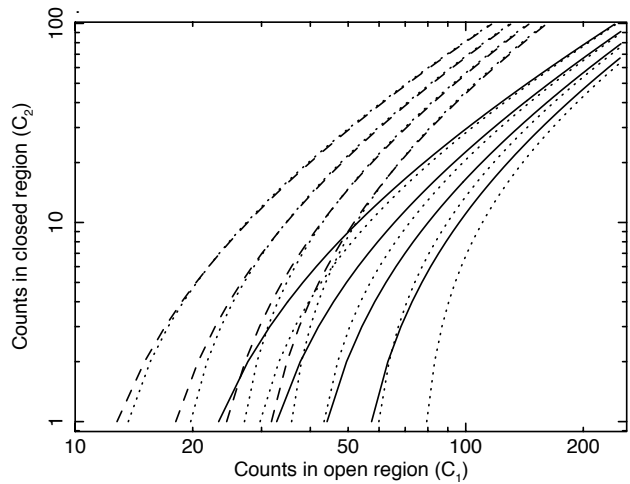} 
  \vspace{-7mm}
   \caption{The numbers of events needed in the region of the detector plane which are exposed to the source ($C_S$) and that shadowed by the mask ($C_B$) needed to achieve a given level of detection significance. The dashed and continuous curves are calculated using the Cash statistic (Equation \ref{cash_equation}) which correctly handles Poisson statistics and are for $f=0.4$ and $f=0.6$ respectively. Dotted curved are based on $\delta\chi^2$ (Equation \ref{chi2_equation}) and so use the approximation of Gaussian statistics.   The curves for each family are  at  levels of 17.3, 26.3, 37.4 and 50.4 (left to right), which for both the Cash statistic and $\chi^2$ correspond to $4\sigma, 5\sigma, 6\sigma$ and $7\sigma$ respectively for one degree of freedom (appropriate if the source position is known).}
   \label{n_photon_plot}
\end{figure}

 If we observe counts $C_S$ and $C_B$ in the exposed and shadowed parts of the detector\footnote{The same symbols are used indiscriminately here for the observed and expected numbers of events as the one is the best estimate of the other.}, the difference in the  Cash likelihood statistic  \cite{\Cash79} between the null (background only) hypothesis and the hypothesis in which a source is present at the supposed position is found to be
 \begin{eqnarray}
{\frac{\delta \mathcal{C}}{ 2}} &  =  &  C_S \ln\left( {\frac{C_S}{f}}\right) + C_B \ln\left( {\frac{C_B} {1-f}}\right)  \nonumber \\
& & -  (C_S+C_B) \ln ( C_S+C_B)    
\label{cash_equation}
\end{eqnarray}
\hfill (assumptions  \ref{a_no_leaks}, \ref{a_perfect_det}, \ref{a_typical},
\ref{a_uniform_bg}).\\
 Figure \ref{n_photon_plot} shows, for two examples of $f$, the combinations of numbers of events which give particular levels of confidence in the detection of a source, calculated according to Equation \ref{cash_equation}. For comparison the corresponding contours can be  calculated on the Gaussian assumption by evaluating the $\chi^2$ parameter
 \begin{eqnarray}
\delta \chi^2& =& {\frac{\left(C_S-f(C_S+C_B)\right)^2}{f(C_S+C_B)}} \nonumber\\
 &  & +{\frac{\left(C_B-(1-f)(C_S+C_B)\right)^2}{(1-f)(C_S+C_B)}}\nonumber \\
 &=&{\frac{\left(C_S-f(C_S+C_B)\right)^2}{f(1-f)(C_S+C_B)}}
\label{chi2_equation}
\end{eqnarray}
\hfill (assumptions  \ref{a_no_leaks}--\ref{a_perfect_det}, \ref{a_typical},
\ref{a_uniform_bg}).\\
In fact  $\delta \chi^2$
in equation \ref{chi2_equation} is just the square of $\hat S/\sigma_i$ from equation  \ref{sn_rel_field}.

Contours of constant $\chi^2$, calculated according to Equation \ref{chi2_equation} are also shown in Figure \ref{n_photon_plot}. 
As the two statistics both follow the $\chi^2(1)$ distribution, a direct comparison can be made. It can be seen that there is relatively little difference between the two except in extreme cases where the {\it total} number of events in the background region number of events is quite low.   The Gaussian assumption can still be good even if number of events {\it per detector element} is small ({even  $\ll 1$), provided the {\it total} number of background events, $C_B$, is more than a dozen or so. For this reason assumption \ref{a_gaussian}   is almost always valid.

\section{Finite detector resolution -- Assumption \ref{a_perfect_det}}

There remains the assumption that the detector has perfect spatial resolution. Unfortunately relaxing this assumption has a major impact. The situation is simplest if we again assume background dominated conditions (assumption \ref{a_bg_dominated}). We consider this case first.
\label{coding_power}
\subsection{Background dominated case}

\label{coding_power_BG}

The case where the mask shadow is recorded by a detector with limited spatial resolution  and where $B_{f,t}\gg S$ is considered by Skinner\cite{\Skinnera95}. It is shown there that in this case the sensitivity relative to that of a reference system with $f=\half$ (Equation \ref{snr_ref_eqn}) is given by the ``coding power",~ $\Delta$, so that
\begin{equation}
   \left({\frac{S}{\sigma_S}}\right) = \Delta \left({\frac{S}{\sigma_S}}\right)_{ref}= \Delta {\frac{S}{2}} {\sqrt {\frac{A}{B_{f,t}}}},
\end{equation}
where 
\begin{equation}
{\frac{\Delta^2}{4}}=  {\frac{1}{n}}  \sum_{\rm i} P_{\rm i}^2   -
      \Bigl( {\frac{1}{n}}  \sum_{\rm i}  P_{\rm i} \Bigr)^2
      \label{delta_eqn}
\end{equation}
\hfill (assumptions  \ref{a_bg_dominated}, \ref{a_gaussian}, \ref{a_typical},
\ref{a_uniform_bg})\\
and where  $P_{\rm i}$ is the response in detector element $i$ to a source at the position under consideration, relative to that for a fully exposed element to the same source (so that $0<P_i<1$). Assumption\ref{a_no_leaks} is not needed as $t_0, t_1$ can be taken into account in calculating the $P_i$. Equation \ref{delta_eqn} shows that $\Delta$ is simply twice  the {\it rms} value of $P_i$. It can be quite different even for two close directions  as the shadows of the edges of mask elements may fall differently with respect to the detector elements. If there is a large number of detector elements and their pitch is not commensurate with that of the mask elements then all relative phases of the two arrays will occur with about the same frequency and any such variations will be small. If there is a simple ratio between the pitches, then an average value of $\Delta$ may be used as a measure of the mean sensitivity, averaged over different sky directions (different shifts of the mask shadow) {\it i.e.} we invoke assumption~\ref{a_typical}.

The concept of coding power is a very useful one. It can be used  to derive Equation \ref{b_gt_s_eqn}, which is the limiting form of Equation \ref{sn_basic_eqn}  when $B_{f,t}\gg S$, or the corresponding  form of Equations \ref{basic_with_t}  (or \ref{sn_rel_field_with_t})  in the same limit. But it also allows quantitative treatment of the loss in sensitivity due to finite detector resolution in any background limited case.  A detector with limited spatial resolution records only a blurred version of the shadow of the mask, as illustrated in Figure \ref{mask_fig} and the {\it rms} deviation of $P$ is consequently reduced.
This approach was used in \cite{\Skinnera95} to obtain an expression for the sensitivity of a telescope  having square element of side $m$ and 
50\% open fraction when the detector has  square pixels of finite size  $d$.  Generalizing the result obtained there  to allow for masks with any open fraction and with imperfect transmission and opacity, one finds that the sensitivity  must be multiplied by a coding power factor
 \begin{eqnarray}
 \Delta & = &    \left( 1-{\frac{d}{ 3 m}}\right) (t_1-t_0) \sqrt{4f(1-f)}    \hspace{8mm} if\:\:   m\ge d   \nonumber \\
              &=& {\frac{m}{d}} \left( {1- {\frac{m}{3d}}}\right)(t_1-t_0)\sqrt{4f(1-f)} \hspace{5mm} if\:\:    m\le d    \nonumber \\
\label{mltd}
 \end{eqnarray}
\hfill (assumptions  
\ref{a_bg_dominated}, \ref{a_gaussian}, \ref{a_typical},
\ref{a_uniform_bg}).\\

Analytic solutions in even more general cases are messy and not very revealing, but numerical calculation of $\Delta$ allows the sensitivity of a proposed or actual system to be estimated. As an example we take the case shown in Figure \ref{mask_fig}, which is relevant to the EXIST project. The shadow of a mask with square elements of side $m$,  supported by a  grid structure with bar width $g$, is  imagined to be recorded using a detector having square pixels of side $d$.  The variation of $P$ along one particular line is illustrated in the case $g<d<m$. 

The calculation can be simplified by noting that the mask pattern can be described as the convolution of a single mask hole, side $m$, with a sparse `bed of nails' (2-d Shah \cite{Bracewell86}) function, pitch $p=m+g$, in which only a fraction $f_e$ of the spikes are present. The response function is then obtained by a further convolution with the form of a  detector element. Use of the convolution theorem allows the Fourier Transform of the response to be obtained and Parseval's theorem then gives its mean square value.
  
Example results  are shown in Figure \ref{coding_loss_fig}.   When the supporting grid is present  there is an important loss in sensitivity unless the detector pixels are very small. In effect the loss is due to an increased fraction of intermediate, `gray', levels because the shadow of the fine grid is poorly resolved. 

In real conditions the shadow of the mask cast by an  off-axis source may not simply be a translation of that for an on axis source. The finite thickness of the mask elements and/or that of a supporting grid, or the partial transparency of the  structure  may  modify the  off-axis response. Grindlay and Hong  \cite{\Grindlayb04} have discussed approaches to some of the problems associated with such complications.

\begin{figure}[tbp] 
   \centering
   \includegraphics[angle=0,width=3.4in]  {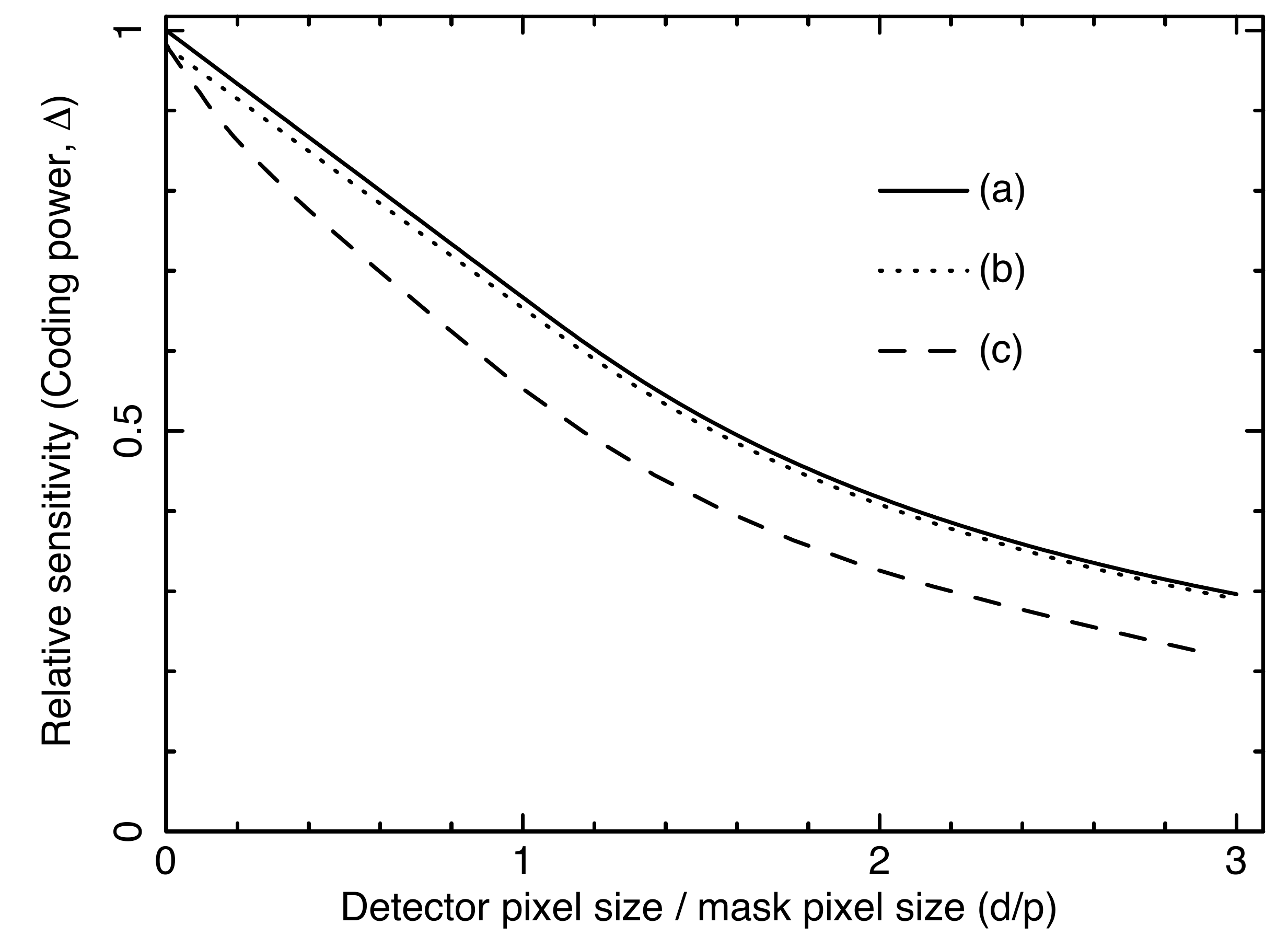} 
   \caption{The loss in sensitivity if the detector spatial resolution is not perfect. The continuous line (a) is for a simple mask with square elements having open fraction $f=0.5$ and a detector with square pixels (Equation \ref{delta_eqn}). The dotted line (b) shows the corresponding curve with $f=0.4$. The dashed line (c) is for a mask in which 50\% of the elements are open but in which a supporting grid like that in Fig. \ref{mask_fig} reduces the transparency to $f=0.4$. If the detector resolution is good (low $d/p$) the grid provides coding and so the curve approaches (b). If the detector resolution is too poor, the grid simply attenuates the flux and reduces the sensitivity. (Assumptions   \ref{a_bg_dominated}--\ref{a_gaussian}, \ref{a_typical}, \ref{a_uniform_bg}).}
   \label{coding_loss_fig}
\end{figure}

\subsection{Finite detector reolution - Background not dominant}

The approach used in \cite{\Skinnera95} and section \ref{coding_power}\ref{coding_power_BG}  for  deriving the expression for the sensitivity considers the problem as equivalent to finding the gradient of the best-fit straight line in a data space relating the observed counts in a detector pixel, $C_i$, to the corresponding $P_i$. In the case treated there, $B_{f,t}\gg S$, so the errors on each point are the same. In the general case the number counts expected in detector pixel $i$, of surface area $a$, is\footnote{Note that the $C_i$ are counts per pixel where $C_S$, $C_B$ were totals for all the pixels of a particular category.}
\begin{equation}
 C_i = a (B_{f,t}+P_i S) \pm \sqrt{ C_i}
\end{equation}
and the best estimate of $S$ can be found as the gradient of the straight line which is a weighted-best-fit to the  points $(C_i,P_i)$. The uncertainty in the gradient is given by
 \begin{equation}
   \sigma_S^2 =  \frac{2}{\Delta_w} \left(  \sum {\frac{1}{ C_i}}\right)^{-1} 
\end{equation}
where $\Delta_w$ is simply the weighted equivalent of  $\Delta$ :
\begin{equation}
{\frac{\Delta_w^2}{4}}=\left( \sum  \frac {P_i^2} { C_i} \right)  /  \left(  \sum \frac{1}{ C_i}  \right)  - \left[ \left( \sum \frac{P_i}{ C_i}  \right) /  \left( \sum \frac{1}{ C_i} \right) \right]^2 
\nonumber
\end{equation}
\hfill (assumptions  \ref{a_gaussian}, \ref{a_snr_src}--\ref{a_uniform_bg}).\\

\subsection{ Optimum open fraction with finite detector resolution}
\label{fopt_detres}

In 't Zand {\sl et al.}  \cite{\intZand94}  have noted how imperfect detector spatial resolution tends to decrease the optimum open fraction. For the case they discussed, that of the Beppo-SAX wide field cameras, they concluded that $f$ in the range 0.25-0.33 was the best choice. However, this conclusion 
 was more due to the fact that the fields simulated contain several sources whose flux led to a high $b$ than to the effects of the detector resolution. 
 
The shift of $f_{opt}$ to lower values due to imperfect detector resolution can nevertheless be important when studying a single strong source.  The arrows in  Figure \ref{fopt_fig} illustrate the effect with $m/d=2$ (shorter arrows) and with $d=m$ (a case discussed below in \S\ref{opt_pos_accuracy}).

\subsection{Poisson Statistics and finite detector resolution}

Finally if the the detector resolution is  finite {\sl and} the number of events are so small that Poisson statistics must be used, the source flux can be obtained by optimizing the Cash likelihood statistic
\begin{equation}
\frac{\mathcal{C}}{2} = - \sum C_i {\ln (P_i S + B_{f,t})}
\end{equation}
and the confidence limits obtained by finding the $S$ for which $\mathcal{C}$ changes by the required amount (with $B_{f,t}$ refitted).
For calculating the confidence with which the null (background only) hypothesis can be rejected one can calculate
\begin{equation}
\frac{\delta \mathcal{C}}{2}= - \sum C_i {\ln (P_i S + B_{f,t})} + 2 n C_i  {\ln (C_i)},
\end{equation}  
\hfill (only assumptions  \ref{a_snr_src}--\ref{a_uniform_bg} necessary)\\
$n$ being the number of detector pixels. Thus although no useful explicit formulae are available in this case, use of this statistic provides a method for dealing with any particular example.


\section{Optimizing source position determination accuracy}
\label{opt_pos_accuracy}

For a given mask to detector separation, dictated perhaps by spacecraft accommodation considerations and/or a  minimum required field of view,  the angular resolution of a coded mask telescope depends on the mask pixel size and also on that of the detector.  Practical issues usually limit how small the pixels of a detector may be made, while the mask design is usually less subject to constraints. If we suppose the detector pixel size to be fixed, the angular resolution will formally continue to improve as the mask pixels are made smaller and smaller, but as was seen above, with low $m/d$  the significance with which sources are detected  will suffer. 

Simulations confirm that a good approximation to the angular resolution is obtained by taking the Pythagorean sum of the 
the angle subtended by a mask pixel at the detector and that subtended by a detector pixel at the mask. Near the centre of the field of view 
\begin{equation}
\delta\theta^2 = (m/l)^2 + (d/l)^2 
\end{equation}
where $l$ is the mask-detector separation. Note that in a case such as that illustrated in Figure \ref{mask_fig} it is the size of the holes ($m$) that is important, not their pitch.

\begin{figure}[tbp] 
   \centering
   \includegraphics[width=3.4in]{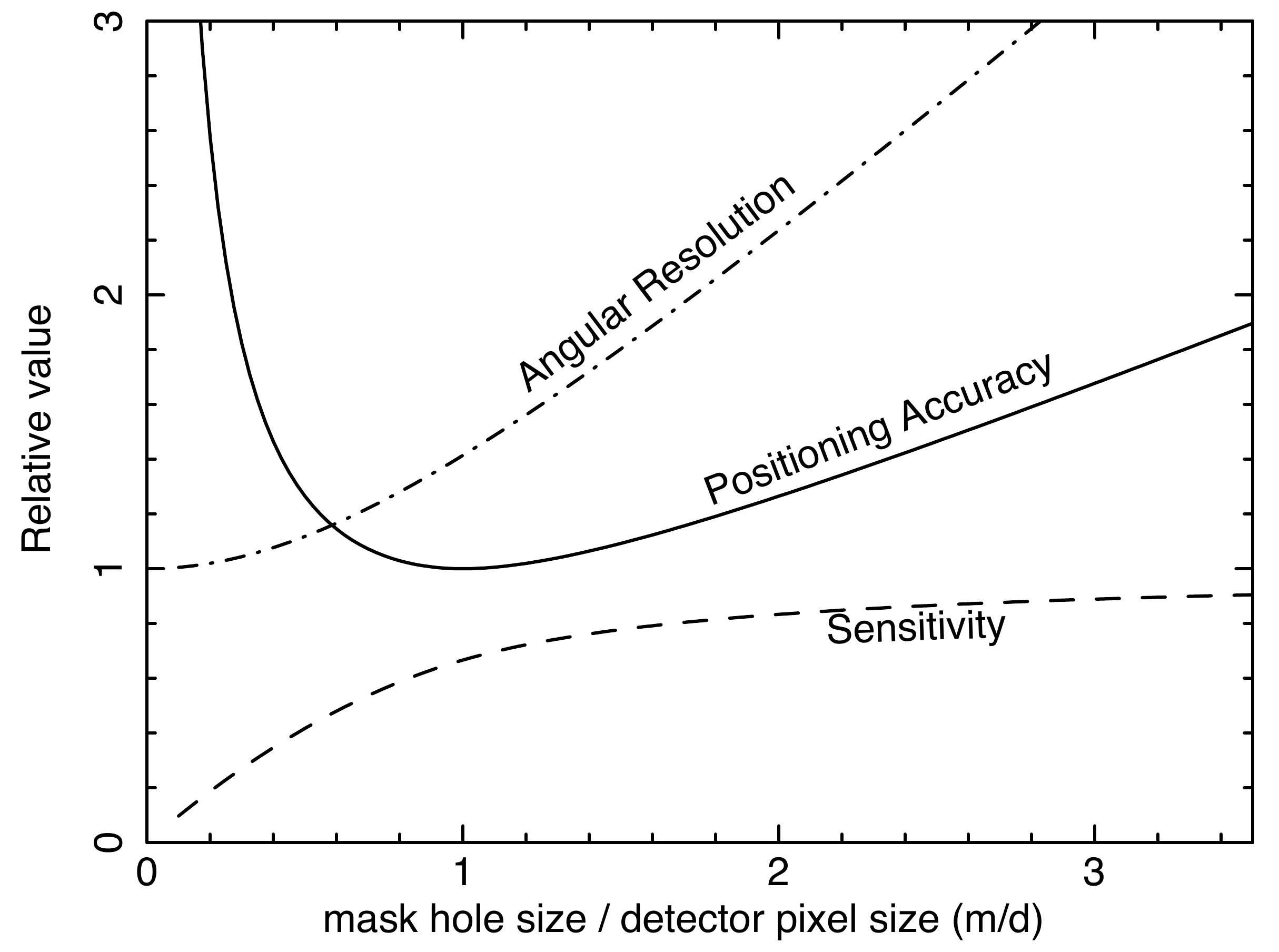} 
   \caption{Variation of relative values of signal-to-noise ratio, angular resolution, and source location accuracy with mask element size for a fixed detector element size.  Assumptions  \ref{a_bg_dominated},  \ref{a_gaussian}, \ref{a_typical}, \ref{a_uniform_bg} are made, in particular noise is assumed to be background dominated ($B_{f,t}\gg S$).}
   \label{optimum_m_fig}
\end{figure}

The accuracy with which a source can be located is better than this by a factor approximately proportional to the signal-to-noise ratio $S/\sigma_S$ of the source\footnote{Sometimes  $S/(\sigma_S-1)$ is used for a better approximation but the differences are small in practical cases.} so
\begin{eqnarray}
\delta\alpha & = &  \left( {\frac{\sigma_S}S}\right)  k \left[(m/l)^2 + (d/l)\right]^{\half}  \nonumber \\
                      & = & \Delta\;   \left({\frac{\sigma_S}S}\right)_{d=0} k \left[(m/l)^2 + (d/l)\right]^{\half},
\end{eqnarray}
where $k$ is a constant of the order of unity which depends on the exact definition of location accuracy. Substituting $\Delta$ from Equation \ref{mltd} one finds that, on the assumptions under which these equations apply (\ref{a_bg_dominated},  \ref{a_gaussian}, \ref{a_typical}, \ref{a_uniform_bg}), the lowest position uncertainty is obtained with $m=d$. This is illustrated in Figure \ref{optimum_m_fig}.

In the design of an instrument the number of objects detectable is likely to also be a consideration.  As the minimum is relatively shallow, choosing a slightly higher $m/d$ will allow additional faint sources to be detected at the expense of only a small loss in positioning accuracy for brighter ones.   The BAT instrument on SWIFT uses $5/4$; a value of $2/1$ is baselined for EXIST.

\section{Conclusions}

The formulae presented above offer insight into the way in which the inevitable uncertainties due to Poisson statistics affect measurements with coded mask telescopes. In some case the differences between a simplified treatment and the more precise one can be quite large. For example with the choice of $m/d=1$, shown in section \S\ref{opt_pos_accuracy} to give the best source location accuracy,  
the sensitivity is worse by a factor $2/3$  than that which would be expected by blind application of a simplified formula. Often in astronomy the number of objects observed depends on the $-3/2$ power of the  detection threshold, so use of the simplified approach would lead to an overestimate of that number by a factor 1.8. 

Although the discussion here has been in terms of  measuring the flux from a particular direction, for example that from a point source, in many cases the results can be applied to extended sources by considering the flux {\sl per angular resolution element} from the source. 

It should be noted that systematic errors due to (uncorrected) variations in the background level across detector plane have not been considered nor have been those due to `ghosts' (sidelobes) of other sources. Thus the results are most directly applicable in the case of short observations of relatively bright sources ({\it e.g.} gamma-ray bursts) or where the design or observation strategy is such as to minimize such errors ({\it e.g.} through use of scanning, combined with a very large number of detector pixels, as in the proposed EXIST black hole finder mission).    

The author wishes to thank Roberto Accorsi, David Band, Jean in 't Zand, Ed Fenimore and Craig Markwardt for helpful discussions.


\end{document}